\documentclass[preprint,bibnotes]{revtex4-1}
\usepackage{amsfonts}
\usepackage{amsmath}
\usepackage{amssymb}
\usepackage{graphicx}

\begin{document}

\title[\\
]{On the theory of the vortex state in the Fulde-Ferrell-Larkin-Ovchinnikov
(FFLO) phase}
\author{Vu Hung Dao}
\affiliation{Laboratoire CRISMAT, CNRS UMR 6508, ENSICAEN, Universit\'e de Caen,
6 Bd Mar\'echal Juin, 14050 Caen Cedex 4, France}

\author{D. Denisov}
\affiliation{Van der Waals-Zeeman Institute, University of Amsterdam, The
Netherlands}

\author{A. Buzdin}
\altaffiliation{also at Institut Universitaire de France, Paris, France} 
\affiliation{Condensed Matter Theory Group, LOMA, UMR\ 5798, University of Bordeaux, 
F-33405 Talence, France}

\author{J. P. Brison}
\affiliation{CEA Grenoble, INAC, SPSMS, F-38054 Grenoble 9, France}


\begin{abstract}
We demonstrate that the vortex state in the Fulde-Ferrell-Larkin-Ovchinnikov
(FFLO) phase may be very different depending on the field orientation
relative to the crystalline axes. We calculate numerically the upper
critical field near the tricritical point taking into account the modulation
of the order parameter along the magnetic field as well as the higher Landau
levels. For s-wave superconductors with the anisotropy described by an
elliptical Fermi surface we propose a general scheme of the analysis of the
angular dependence of upper critical field at all temperatures on the
basis of the exact solution for the order parameter. Our results show that
the transitions (with tilting magnetic field) between different types of
mixed states may be a salient feature of the FFLO phase. Moreover we discuss
the reasons for the first-order phase transition into the FFLO state in the case of
CeCoIn$_{5}$ compound.
\end{abstract}

\date{}
\maketitle

\preprint{}

\volumeyear{} \volumenumber{} \issuenumber{} \eid{}

\received[Received text]{} \revised[Revised text]{} 
\accepted[Accepted
text]{} \published[Published text]{} \startpage{1} \endpage{}

\section{Introduction}

\bigskip Recent experimental studies of the superconducting state of CeCoIn$%
_{5\text{ }}$(see \cite{MatsudaShimahara07} and references cited therein)
provided evidences in favor of the Fulde-Ferrell-Larkin-Ovchinnikov (FFLO)
phase existence in the high magnetic field region of the superconducting
phase diagram. Originally \cite{FuldeFerrel, LarkinOvchinnikov} the
nonuniform FFLO state has been predicted to exist in superconductors when
the magnetic field is acting on the electron spins only (the case of the
paramagnetic effect). Usually it is an orbital effect which is the most
important and this makes difficult the experimental observation of the FFLO
phase. Moreover the superconductor must be in the clean limit because the
electron scattering is detrimental for the FFLO phase \cite{Aslamazov}. The
orbital effect may be weakened in heavy fermion superconductors or in
quasi-2D superconductors when magnetic field is applied parallel to the
superconducting planes. That is why in addition to the heavy fermion
superconductor CeCoIn$_{5\text{ }}$, quasi-one and quasi-two-dimensional
organic superconductors are considered as good candidates for the FFLO
phase realization \cite{buzdin00},\cite{lebe00}. Recently evidences of
the FFLO state have been revealed in organic quasi-2D superconductors $%
\lambda $-(BETS)$_{2}$ FeCl$_{4}$ \cite{Uji01} and $\kappa $-(BEDT-TTFS)$%
_{2} $ Cu(NCS)$_{2}$ \cite{Izawa}.

In the framework of an isotropic model with s-wave pairing the critical
field for the FFLO phase in the presence of the orbital effect has been
calculated by Gruenberg and Gunther \cite{GruenbergGunther}. They
demonstrated that the FFLO state may exist if the ratio of pure orbital
effect $H_{c2}^{orb}(0)$ and pure paramagnetic limit $H_{p}(0)$ is larger
than 1.28, i.e. the Maki parameter $\alpha _{M}=\sqrt{2}%
H_{c2}^{orb}(0)/H_{p}(0)$ is larger 1.8. The pure paramagnetic limit at $T=0$
can be estimated as $H_{p}(0)=\Delta _{0}/\sqrt{2}\mu _{B}$, where $\Delta
_{0}$ is the BCS gap at $T=0$ and $\mu _{B}$ is the Bohr magneton \cite%
{SaintJamesSarma}. In \cite{GruenbergGunther} the exact solution for the
order parameter was described by an FFLO modulation along magnetic field and the zero
Landau level function for the coordinates in the perpendicular plane. Further
analysis \cite{BuzdinBrison} revealed that the higher Landau level solutions
(LLS$)$ become relevant for large values of Maki parameter $\alpha _{M}>9$
and the $H_{c2}(T)$ curve may present regions described by different LLS.
These results obtained for an isotropic model are readily generalized for
the case where the electron spectrum anisotropy is described by an elliptic
Fermi surface \cite{Brison95} . In such a case the Maki parameter
becomes angular dependent and the transitions between different LLS may
occur with a change of orientation of the magnetic field.

It happens that for an adequate description of the FFLO state in real
compounds, the form of the Fermi surface as well as the type of the
superconducting pairing play a very important role because they determine
the direction of the FFLO modulation. This circumstance has been
demonstrated \cite{Denisov,BuzdinMatsuda07} in the framework of a general
phenomenological approach based on the modified Ginzburg-Landau (MGL)
functional \cite{BuzdinKachKachi97}. This approach is adequate near the
tricritical point (TCP) in the field-temperature phase diagram. At the TCP
the three transition lines meet: the lines separating the normal metal, the
uniform superconducting state and the FFLO state. Near the TCP the wave
vector of FFLO modulation is small and this situation may be described by
the MGL functional. For the case when the deviation of the Fermi surface
from the elliptical form is small the method \cite{Denisov} permits to
calculate the critical field corresponding to different LLS.

(Section~\ref{sec:2}) Unfortunately the approach \cite{Denisov} is limited to  weak
deviations from the elliptic Fermi surface. In Section~\ref{sec:2} we develop a
numerical method for the calculation of the upper critical field applicable to
any cases, using a general form of the solution for the order parameter
as a superposition of the different LLS. Note that the single LLS is an
exact solution for the order parameter only for isotropic or quasi-isotropic
(elliptic Fermi surface) cases. Otherwise the order parameter is described
by an infinite serie of the LLS. However there is usually some dominating
LL $n_{0}$ and the amplitudes of other LL rapidly decrease with an
increase of $\left\vert n-n_{0}\right\vert $. Our analysis qualitatively
confirm the conclusions of \cite{Denisov} and reveal the transitions between
the FFLO states with different dominating $n_{0}$. The obtained results
demonstrate that the FFLO state, depending on superconductor parameters 
and/or magnetic field orientation, may take the form of the higher
LLS with or without a modulation along the magnetic field. The transitions
betwen these states result in a very rich dependence of the transition
temperature on the magnetic field orientation.

(Section~\ref{sec:3}) The approach of Section~\ref{sec:2} based on MGL is adequate near the TCP.
On the other hand the case of anisotropic superconductors with elliptic
Fermi surface may be treated exactly at all temperatures. In section~\ref{sec:3} we
use the scaling transformation \cite{Brison95} to obtain the solutions for
the higher LLS. As an illustration we consider quasi-1D and quasi-2D
superconductors. Note that the higher LLS in quasi-2D superconductors with
the in-plane orientation of the magnetic field were studied by Shimahara 
\cite{Shmahara2009}.

(Section~\ref{sec:4}) The experiments on CeCoIn$_5$ show that at low
temperature the superconducting transition becomes of the first order \cite{Izawa2}. 
In Section~\ref{sec:4} we argue that this may be explained by the magnetism
associated with cerium cites and its interaction with the superconducting
subsystem. In the framework of the proposed model in the mixed state the cerium
polarization should strongly increase in the cores of vortices. This
mechanism could contribute to the anomalously large form factor of the
vortex structure observed in CeCoIn$_{5}$ at low temperature \cite{Bianchi}, 
\cite{White}.

\bigskip

\section{Formation of the different FFLO states under
the influence of the orbital effect} \label{sec:2}

In this section we provide a general numerical approach for the calculation
of the upper critical field of the FFLO state. Keeping in mind CeCoIn$_5$
 we will consider the case of the tetragonal symmetry. Usual GL
functional contains only the first derivatives of the order parameter and
therefore it may be transformed by simple scaling transformation of the
coordinates to the isotropic form (with the corresponding renormalization of
the magnetic field). That is why in the framework of the GL approach we may
easily obtain the exact solution of the $H_{c2}$ problem for any anisotropic
superconductor - the order parameter is decribed by the n=0 LL function \cite
{AbriskovBook88}. It is possible to decribe the FFLO state near the TCP
point by MGL which takes into accound the higher derivatives of the order
parameter. In contrast to the GL functional the MGL functional can not be
reduced to the isotropic case by scaling the coordinates (this is possible
only for s-wave superconductivity with an elliptic Fermi
surface \cite{Brison95}). Therefore in the most general case, after performing the
scaling transformation which makes the part with the first derivatives
isotropic, we have the following modified Ginzburg-Landau functional: 
\begin{equation}
\mathcal{F}=\Psi ^{\ast }\Big(\alpha +\hat{L}\Big)\Psi  \label{eq:Ftetra}
\end{equation}
with $\alpha (H,T)=\alpha _{0}(T-T_{cu}(H))$ where $T_{cu}(H)$ is the
transition temperature into the uniform superconducting state. The
differential operator has the general expression 
\begin{equation}
\hat{L}=-g\sum_{j}\Pi _{j}^{2}+\gamma \Big(\sum_{j}\Pi _{j}^{2}\Big)%
^{2}+\varepsilon _{z}\Pi _{z}^{4}+\frac{\varepsilon _{x}}{2}\{\Pi
_{x}^{2},\Pi _{y}^{2}\}+\frac{\tilde{\varepsilon}}{2}\Big(\{\Pi _{x}^{2},\Pi
_{z}^{2}\}+\{\Pi _{y}^{2},\Pi _{z}^{2}\}\Big),
\end{equation}%
where $\Pi _{j}=-i\partial _{j}+2\pi A_{j}/\Phi _{0}$ (with $j=x,y,z$), $%
\Phi _{0}$ is the flux quantum, and the anti-commutator $\{O_{1},O_{2}\}%
\equiv O_{1}O_{2}+O_{2}O_{1}$. Here the $z$ axis is perpendicular to the basal plane.
To simplify the discussion, we assume that the effective mass is isotropic 
(taking account of its anisotropy is detailed in Appendix). 
For the appearance of the FFLO state $g$ must be positive.
The coefficients $\varepsilon _{z}$, $\varepsilon _{x}$, $\tilde{\varepsilon}$
describe the deviation of the actual Fermi surface from the elliptic one
and/or \ the pairing different from s-wave type. In contrast with previous
work \cite{Denisov} they are not assumed to be small.

The transition temperature is given by $T_{c}(H)=T_{cu}(H)-\lambda _{\mathrm{%
min}}/\alpha _{0}$ where $\lambda _{\mathrm{min}}$ is the smallest
eigenvalue of the operator $\hat{L}$. Within the coordinate frame $%
(x^{\prime },y^{\prime },z^{\prime })$ with the $z^{\prime }$ axis pointing
in the direction of the field, the eigenfunctions of $\hat{L}$ can be looked
for in the form $\Psi =\exp (iq_{z}z^{\prime })\phi (x^{\prime },y^{\prime
}) $ where $q_{z}$ is the FFLO modulation vector along the field direction.
In the abscence of anisotropic fourth-order terms, $\phi $ can be found
exactly. It is one of the Landau levels $\varphi _{n}$ defined in the $(
x^{\prime},y^{\prime})$ plane. The eigenvalues are then 
\begin{equation}
\lambda ^{\mathrm{iso}}(q_{z},n)=\gamma \big[\big((2n+1)\xi
_{H}^{-2}+q_{z}^{2}-q_{0}^{2}\big)^{2}-q_{0}^{4}\big]  \label{eq:lambda_iso}
\end{equation}%
where $n$ is a positive integer, the magnetic length 
\begin{equation}
\xi _{H}\equiv \sqrt{\frac{\Phi _{0}}{2\pi H}},
\end{equation}%
and the modulation vector 
\begin{equation}
q_{0}\equiv \sqrt{\frac{g}{2\gamma }}.
\end{equation}%
In this case, $\lambda _{\mathrm{min}}=-\gamma q_{0}^{4}=-g^{2}/4\gamma $
with a degeneracy of solutions $(q_{z},n)$ which is lifted when anisotropy
is present. In the general case, we diagonalize $\hat{L}$ in the subspace of
functions $\varphi _{q_{z},n}\equiv \exp (iq_{z}z^{\prime })\varphi
_{n}(x^{\prime },y^{\prime })$ (see details in Appendix) in order to find
the smallest eigenvalue $\lambda (q_{z})$ which is then minimized with
respect to $q_{z}$ to get $\lambda _{\mathrm{min}}$.

\begin{figure}[tbp]
\begin{center}
\scalebox{0.92}{
 \includegraphics*{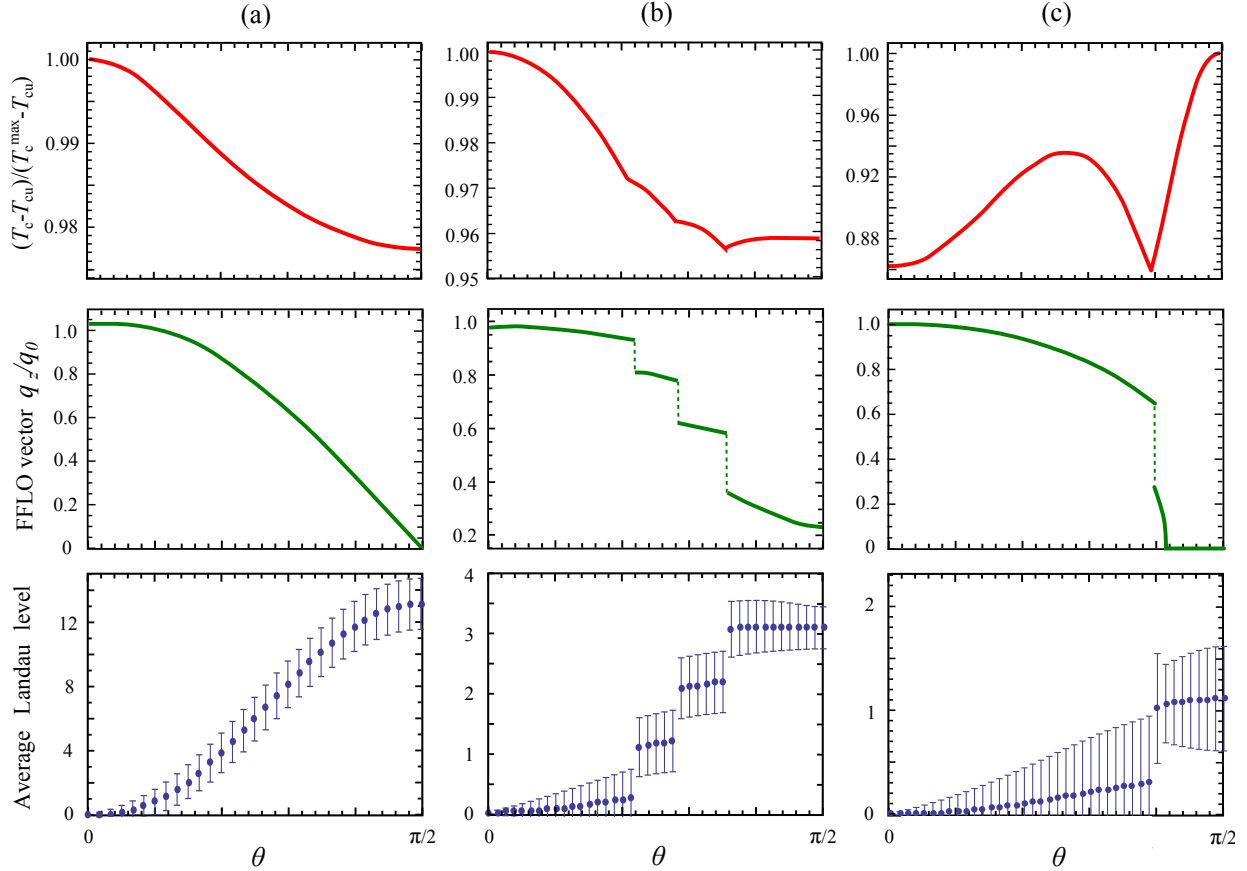}}
\end{center}
\caption{Angle dependence of the transition temperature and of the
corresponding FFLO state for parameters $\protect\varepsilon_x = \tilde{%
\protect\varepsilon}=0$, with (a) $\protect\varepsilon_z= -0.1 \protect%
\gamma $, $g=50 \protect\gamma \protect\xi_H^{-2}$, (b) $\protect\varepsilon%
_z= -0.1 \protect\gamma$, $g=15 \protect\gamma \protect\xi_H^{-2}$, and (c) $%
\protect\varepsilon_z= -0.5 \protect\gamma$, $g=4 \protect\gamma \protect\xi%
_H^{-2}$.  The vertical bars in the bottom plots show the mean square deviation
 of the Landau levels composing the state from the average LL.}
\label{fig:angle-a}
\end{figure}

\begin{figure}[tbp]
\begin{center}
\scalebox{0.92}{
 \includegraphics*{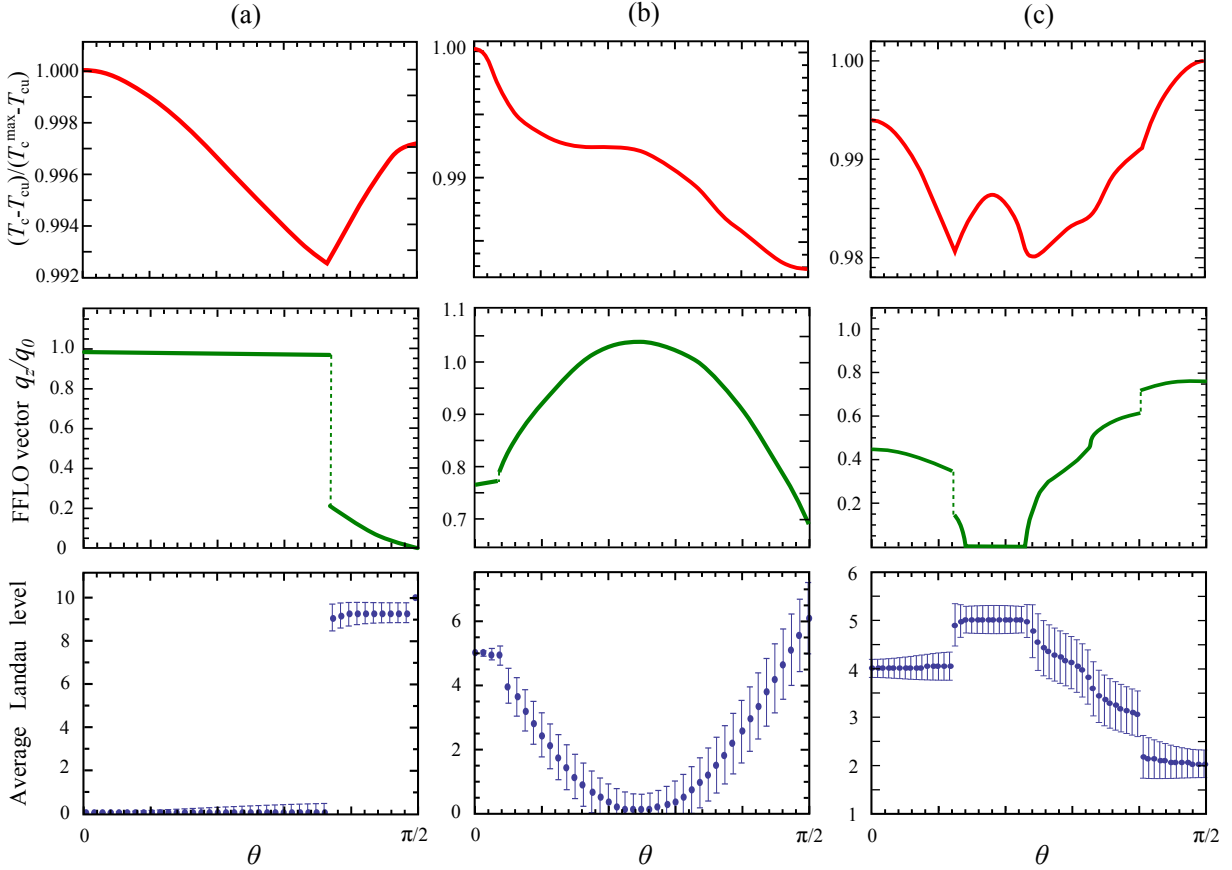}}
\end{center}
\caption{Angle dependence of the transition temperature and of the
corresponding FFLO state for parameters (a) $\tilde{\protect\varepsilon}=%
\protect\varepsilon_z =0$, $\protect\varepsilon_x= 0.5 \protect\gamma$, $%
g=40 \protect\gamma \protect\xi_H^{-2}$, (b) $\protect\varepsilon_x = 
\protect\varepsilon_z=0$, $\tilde{\protect\varepsilon}= -0.5 \protect\gamma$%
, $g=40 \protect\gamma \protect\xi_H^{-2}$, and (c) $\tilde{\protect%
\varepsilon}=\protect\varepsilon_x =-0.3 \gamma$, $\protect\varepsilon_z= 0.2 
\protect\gamma$, $g=20 \protect\gamma \protect\xi_H^{-2}$. The vertical bars
in the bottom plots show the mean square deviation of the LL composing the state  
from the average LL.}
\label{fig:angle-b}
\end{figure}

\begin{figure}[tbp]
\begin{center}
\scalebox{1}{
 \includegraphics*{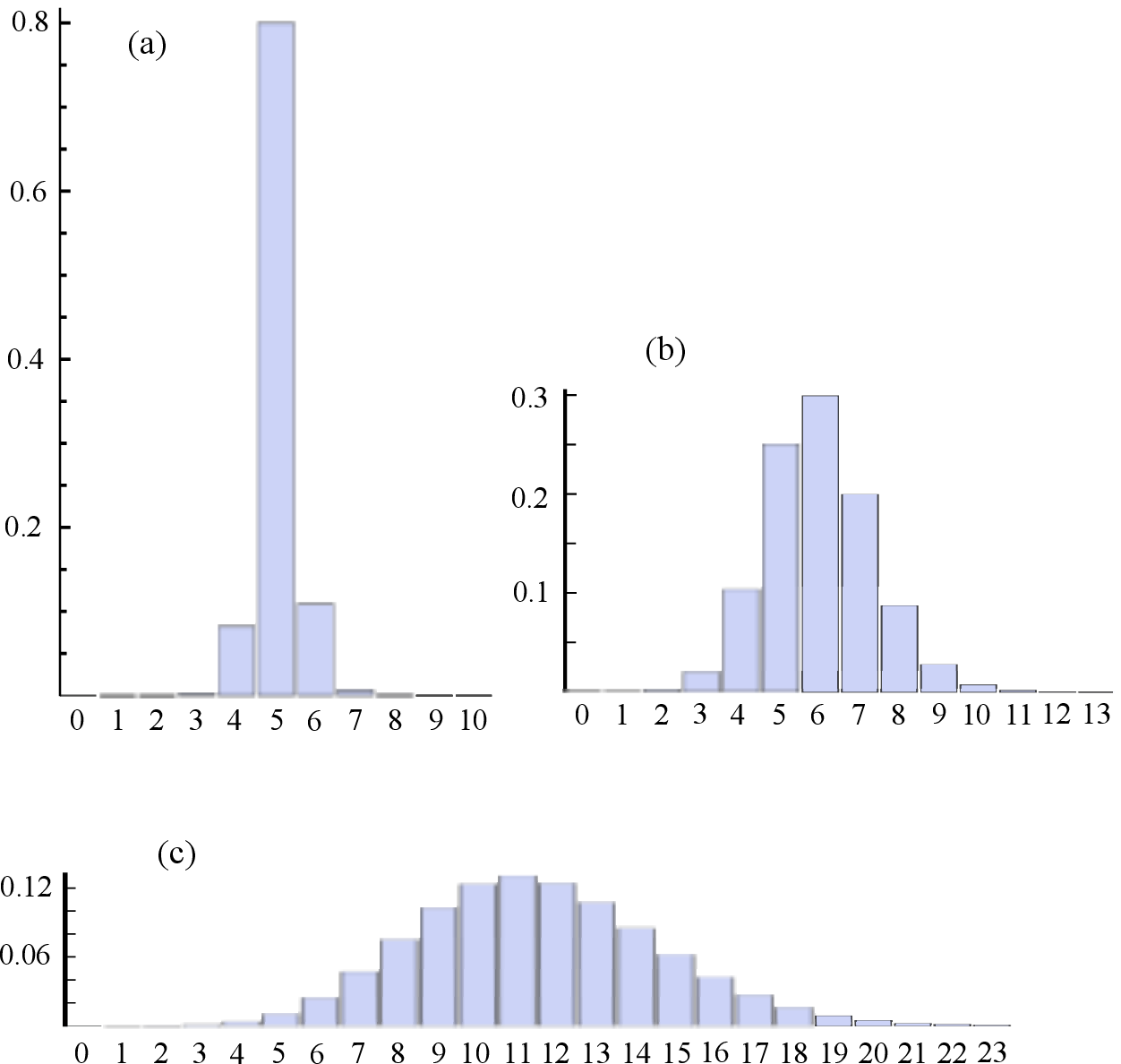}}
\end{center}
\caption{Weight $|c_n|^2$ of the n-th Landau level $\protect\varphi_n$ in
the expansion of the FFLO state $\Psi = \exp{(i q_z z)} \sum_n c_n \protect%
\varphi_{n} $ at the field angle $\protect\theta=\protect\pi/4$, for
parameters $\protect\varepsilon_x = \tilde{\protect\varepsilon}=0$, $g= 50 
\protect\gamma \protect\xi_H^{-2}$, with (a) $\protect\varepsilon_z= -0.01 
\protect\gamma$, (b) $\protect\varepsilon_z= -0.1 \protect\gamma$, and (c) $%
\protect\varepsilon_z= -0.5 \protect\gamma$.}
\label{fig:LL_exp}
\end{figure}

A previous work~\cite{Denisov} showed that due to the effect of anisotropy
three different types of solution for the FFLO state can be realized: (a)
the maximum modulation occurs along the magnetic field with the zero Landau level
state, (b) both modulation and higher Landau level state, (c) the highest
possible Landau level and no modulation along the field (or a modulation with
a very small wave-vector). Moreover due to the specific form of the Fermi
surface a variation of magnetic field orientation may provoke transitions
between states with different Landau levels. However a single-level
approximation was used to get analytical results for these solutions. Due to
this approximation the analytical results were valid for magnetic field
higher than $H\gg \Phi _{0}\frac{g}{\gamma }\sqrt{\frac{\varepsilon }{\gamma 
}}$. In the present work we show using numerical calculations that taking
into consideration the full set of Landau levels, the results qualitatively 
remain true for arbitrary values of the magnetic field $H$.

We calculate the transition temperature and the corresponding FFLO state
when the magnetic field is applied in the $xz$ plane. Typical results are
displayed in Fig.~\ref{fig:angle-a} where we have set $\varepsilon _{x}=%
\tilde{\varepsilon}=0$. The form of the FFLO solution depends only on the
parameter ratios $\varepsilon _{z}/\gamma $ and $\xi _{H}\sqrt{\frac{g}{%
2\gamma }}$ (see e.g. expression~(\ref{eq:Lmn}) of the operator $\hat{L}$ in
the basis of LL). As illustrated in Fig.~\ref{fig:angle-a}(a), the FFLO
state can appear with the maximum modulation vector $q_z=q_{0}$ and the $n=0$ LL when the
field is along the $z$ axis. In contrast, for $H$ along the $x$ axis, there
is no longitudinal modulation and the solution is composed by higher
LL, which results in transverse modulations of the order
parameter. When the field is rotated from one axis to the other, the state
is transformed with a continuous variation of the FFLO modulation and a
smooth evolution of its expansion over the LL. However for smaller values of 
$\xi _{H}q_{0}$ or $\varepsilon _{z}/\gamma $ the variation with the field
orientation can be discontinuous with jumps of the FFLO modulation vector
(see Figs.~\ref{fig:angle-a}(b) and \ref{fig:angle-a}(c)). The sharp transitions 
are manifested by bumps and kinks in the angle dependence of the transition temperature.
Fig.~\ref{fig:angle-b}(a) shows that, for other anisotropy parameters, the jump can
occur between states separated by more than one LL. As expected
from the condition of single-level approximation $H\gg \Phi _{0}\frac{g}{%
\gamma }\sqrt{\frac{\varepsilon }{\gamma }}$ or equivalently $\xi
_{H}^{2}q_{0}^{2}\sqrt{\frac{\varepsilon }{\gamma }}\ll 1$, the number of LL
that contribute significantly in the expansion of the FFLO state increases
with the inverse of the field and/or the anisotropy (see Fig.~\ref{fig:LL_exp}). 
The broadening of the expansion over the LL ends up in supressing the discontinuities. 
In addition, as illustrated in Figs.~\ref{fig:angle-b}(b) and ~\ref{fig:angle-b}%
(c), the FFLO modulation and the transition temperature can vary
non-monotonuously with the field angle. It is interesting to note that at the
angles when the average LL is maximum the wave-vector of
modulation is minimal (and vice versa) and it can even drop to zero (see
Fig.~\ref{fig:angle-b}(c)). At these regions the FFLO state corresponds to
the highest LL states only. The experimental observation of such a
non-trivial angular dependence of $T_{c}$ would be a strong evidence of the
FFLO state.

\section{Higher Landau level states in the framework of the model of
effective mass anisotropy} \label{sec:3}

In this section we demonstrate how the higher LLS naturally appear in the
exactly solvable model of the FFLO transition in a framework of anisotropic
effective mass model. As it has been already noted in the case of the pure
paramagnetic limit this model is reduced by a scaling transformation to the
isotropic one with an arbitrary direction of the FFLO modulation \cite{Brison95}.
 In the presence of the orbital effect the situation is different
and we consider here the uniaxial anisotropy (note that our results are
readily generalized to the arbitrary anisotropy case). We are interested by
a part of the Hamiltonian depending on the magnetic field $\mathcal{H}_{orb}$
+$\mathcal{H}_{Pauli}$ with

\begin{equation}
\mathcal{H}_{orb}=-\frac{1}{2m}\left( \frac{\partial}{\partial x}-\frac {ie}{%
c}yH\cos\theta\right) ^{2}-\frac{1}{2m}\left( \frac{\partial}{\partial y}%
\right) ^{2}-\frac{1}{2m_{z}}\left( \frac{\partial}{\partial z}-\frac {ie}{c}%
yH\sin\theta\right) ^{2},
\end{equation}

\begin{equation}
\mathcal{H}_{Pauli}=\mu_{B}H\sigma_{z},
\end{equation}
where we consider the effective mass $m_{x}=m_{y}=m$ and the magnetic field $%
\mathbf{H}$ is in the $xz$ plane making an angle $\theta $ with the $z$ axis.
 Our treatment can be readily generalized to the case of an anisotropic $%
g$ factor \cite{Brison95}. The gauge of the vector potential $\mathbf{A}$ is
chosen as \ $A_{x}=yH\cos \theta ,$ $A_{y}=0,$ $A_{z}=yH\sin \theta $ and
the spin quantization axis is along the magnetic field.

Performing the scaling transformation $z=z'\sqrt{\frac{m}{m_{z}}}$ the
orbital part becomes \cite{Brison95}
\begin{equation}
\mathcal{H}_{orb}=-\frac{1}{2m}\left( \frac{\partial}{\partial x}-\frac {ie}{%
c}yH\cos\theta\right) ^{2}-\frac{1}{2m}\left( \frac{\partial}{\partial y}%
\right) ^{2}-\frac{1}{2m}\left( \frac{\partial}{\partial z'}-\frac {ie}{c}%
yH\sqrt{\frac{m}{m_{z}}}\sin\theta\right) ^{2},
\end{equation}
i.e. it corresponds to the isotropic metal with an effective mass $%
m $ at the orbital magnetic field $\widetilde{H}=H\sqrt{\cos^{2}\theta+%
\frac {m}{m_{z}}\sin^{2}\theta}$\ (\ $\widetilde{H}_{z}=H_{z}$ and $%
\widetilde {H}_{x}=H_{x}\sqrt{\frac{m}{m_{z}}}$ ). The Pauli contribution
may be written as 
\begin{equation}
\mathcal{H}_{Pauli}=\mu_{B}H\sigma_{z}=\frac{\mu_{B}\widetilde{H}\sigma_{z}}{%
\sqrt{\cos^{2}\theta+\frac{m}{m_{z}}\sin^{2}\theta}}=\widetilde{\mu}_{B}%
\widetilde{H}\sigma_{z},
\end{equation}
with the angular dependent effective Bohr magneton $\widetilde{\mu}%
_{B}\left( \theta\right) =\mu_{B}/\sqrt{\cos^{2}\theta+\frac{m}{m_{z}}%
\sin^{2}\theta}.$

In fact we have reduced the problem of the FFLO critical field calculation
to that of the isotropic model with the field $\widetilde{H}$ and the effective
Bohr magneton $\widetilde{\mu}_{B}\left( \theta\right).$ The corresponding
Maki parameter is $\alpha_{M}=\sqrt{2}H_{c2}^{orb}(0)/H_{p}(0)$ with, in our case, 
$H_{c2}^{orb}(0)$ that is determined by the effective mass $m$ and then is the
pure orbital field along the $z$ axis, while $H_{p}(0)=\frac{\Delta_{0}}{\sqrt {%
2}\widetilde{\mu}_{B}\left( \theta\right) }$. So the Maki parameter becomes
angular dependent 
\begin{equation}
\alpha_{M}\left( \theta\right) =\frac{2\mu_{B}H_{c2}^{z\text{ }orb}(0)}{%
\Delta_{0}\sqrt{\cos^{2}\theta+\frac{m}{m_{z}}\sin^{2}\theta}}=0.54\frac{K}{T%
}\left( \frac{dH_{c2}^{{}}(\theta)}{dT}\right) _{T=T_{c0}}.
\end{equation}
Remarkably in the later expression for $\alpha_{M}\left( \theta\right) $
enters only the slope of $H_{c2}$ at the same angle $\theta$. As it was
demonstrated in \cite{BuzdinBrison} for large values of the Maki parameter, 
$\alpha_{M}>9$, the critical FFLO field at low temperature is determined by
higher LLS. In the case of a large quasi-2D anisotropy $\frac{m_{z}}{m} \gg 1$ this
situation is realized when the Maki parameter is strongly increased for a
field orientation near the $xy$ plane. On the contrary for the quasi-1D
anisotropy $\frac{m_{z}}{m}\ll 1$ the Maki parameter is maximum for the field
orientated along the $z$ axis.

The critical field may be numerically calculated from the formula 
\cite{BuzdinBrison}
\begin{equation}
\ln\left( \frac{T}{T_{co}}\right) =\frac{T}{T_{co}}2\pi\operatorname{Re}\underset{%
\omega_{n}>0}{\sum }\left[ \left( -1\right) ^{N}\int\frac{\beta L_{N}\left(
2\beta y\right) }{\sqrt{\widetilde{Q}^{2}+y}}\tan^{-1}\left( \frac{T_{co}}{%
\omega_{n}+i\mu_{B}H}\right) e^{-\beta y}dy-\frac{T_{co}}{\omega_{n}}\right]
\end{equation}
where $T_{co}$ is the (zero field) critical temperature, $\omega_{n}=\pi T\left( 2n+1\right)$
 are the Matsubara frequencies, $L_{N}$ are Laguerre polynomials, and
\begin{equation}
\beta=\frac{T_{co}}{H\left( \theta\right) }\frac{7\zeta(3) }{%
12\pi^{2}}\left( \frac{dH_{c2}^{{}}(\theta)}{dT}\right) _{T=T_{c0}}.
\end{equation}
The LL number $N$ and the dimensionless vector of the FFLO modulation $%
\widetilde{Q}=\hslash v_{F}Q/2T_{co}$ are chosen in a way to give the
maximum critical field $H\left( \theta\right)$.

\begin{figure}[tbp]
\begin{center}
\scalebox{0.12}{
 \includegraphics*{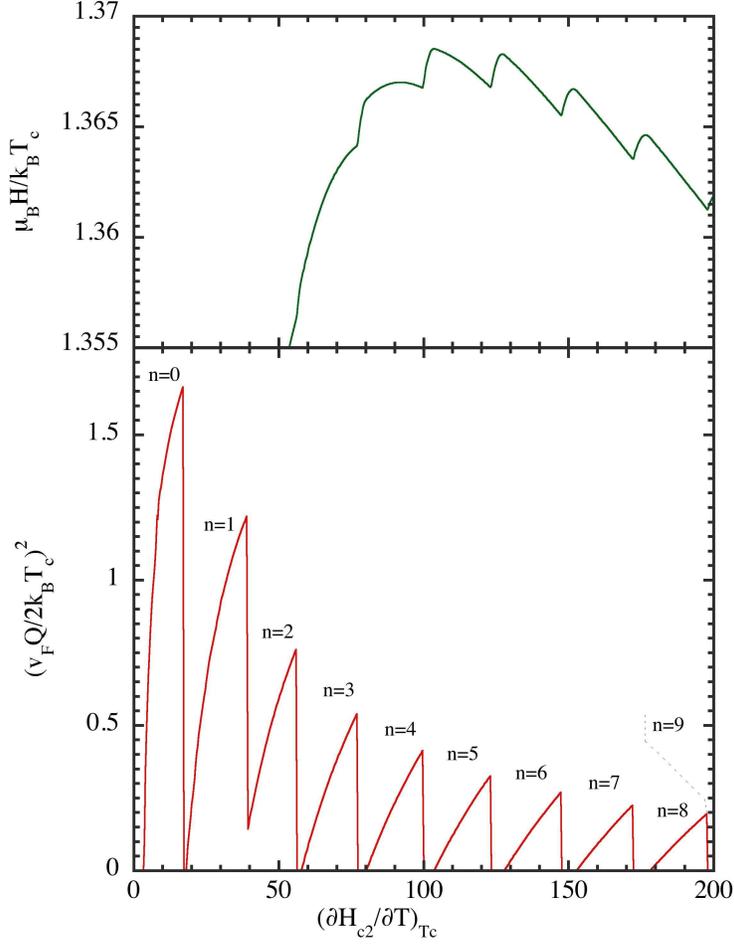}}
\end{center}
\caption{Upper panel: zero temperature critical field as a
function of the initial slope $\left( \frac{dH_{c2}^{{}}}{dT}\right)
_{T=T_{c0}}.$ The transitions between the higher LLS are clearly seen. 
Lower panel: the FFLO modulation vector.}
\label{fig:III-1}
\end{figure}

\begin{figure}[tbp]
\begin{center}
\scalebox{0.15}{
 \includegraphics*{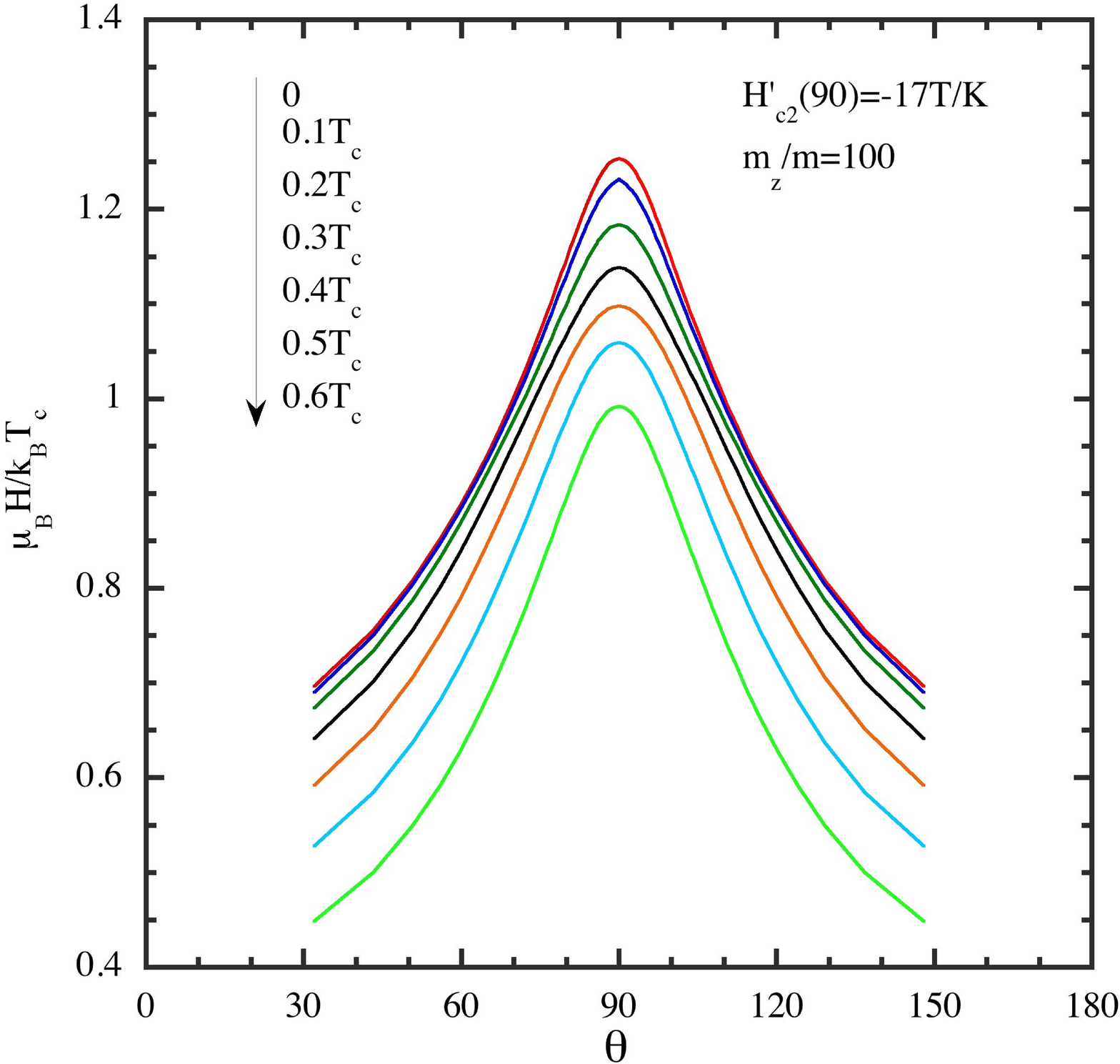}}
\end{center}
\caption{The angular dependence of the upper critical field $
H_{c2}(\theta )$ at different temperature for the initial slope $-\left( 
\frac{dH_{c2}^{{}}(90^{\circ})}{dT}\right) _{T=T_{c0}}=17\left( T/K\right)$.
 This case corresponds to the $n=0$ LL state.}
\label{fig:III-2}
\end{figure}

The calculated values of the upper critical field at $T=0$ $K$ as a
function of the critical field slope at $T=T_{co}$\ are presented in Fig.~\ref{fig:III-1}.
The LLS with $n>0$ appear at $-\left( \frac{dH_{c2}^{{}}(\theta )}{dT}%
\right) _{T=T_{c0}}>18\left( T/K\right) $. We see that with an increase of
the slope the Landau level number $n$ increases, while the FFLO modulation
vector drops. For some slopes it occurs to be zero, and then the FFLO state
is purely higher LLS. In the upper panel of Fig.~\ref{fig:III-1} we observe the
non-monotonous behavior of the upper critical field as a function of the
slope (or the anisotropy $\frac{m_{z}}{m}$). With the increase of the slope
the orbital effect is switched off and we approach the pure paramagnetic
limit for the 3D case. However at $T=0$ $K$ the transition into FFLO state
is a first order transition \cite{Houzet} and then the calculated upper
critical field should be the overcooling field of the normal phase.

\begin{figure}[tbp]
\begin{center}
\scalebox{0.15}{
 \includegraphics*{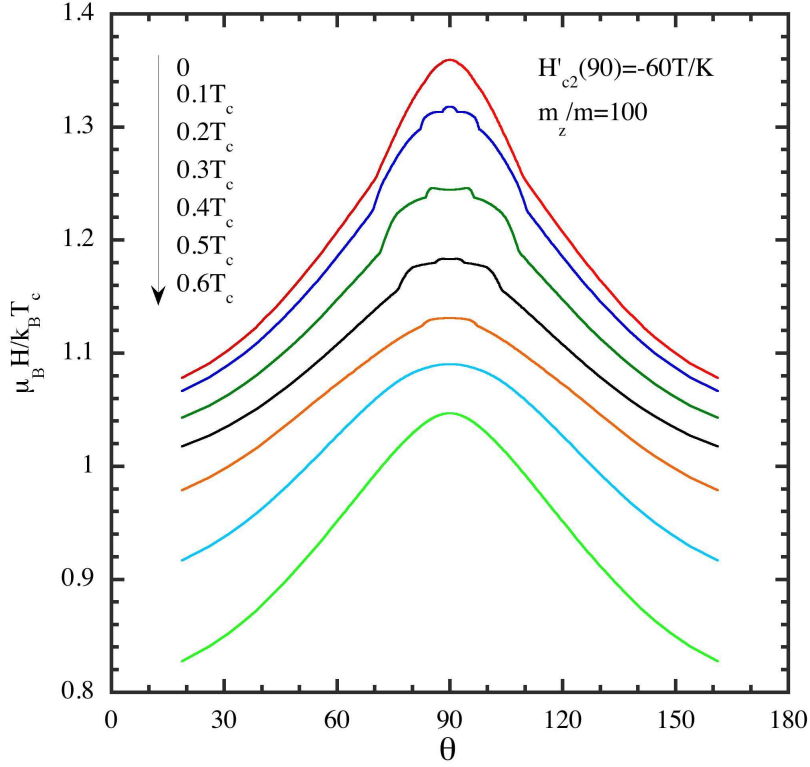}}
\end{center}
\caption{The angular dependence of the upper critical field $%
H_{c2}(\theta )$ at different temperature for the initial slope $-\left( 
\frac{dH_{c2}^{{}}(90%
%TCIMACRO{\U{b0}}%
%BeginExpansion
{{}^\circ}%
%EndExpansion
)}{dT}\right) _{T=T_{c0}}=60\left( T/K\right) .$ At low temperature the
transitions between the different LLS \ are responsible for the peculiar
form of $H_{c2}(\theta )$ dependence.}
\label{fig:III-3}
\end{figure}

Large values of the Maki parameter suitable for the observation of these higher LLS states are mainly expected in strongly (quasi 2D or quasi 1D) anisotropic systems. In such systems, the formation of the higher LLS may be clearly observed on the angular
dependence of the critical field, which will reproduce the dependence on the initial slope displayed on Fig.~\ref{fig:III-1} . In Fig.~\ref{fig:III-2}, such an angular dependence is
presented for a maximum slope $-\left( \frac{dH_{c2}^{{}}(90%
%TCIMACRO{\U{b0}}%
%BeginExpansion
{{}^\circ}%
%EndExpansion
)}{dT}\right) _{T=T_{c0}}=17\left( T/K\right) ,$ with a ratio of effective masses  $\frac{m_{z}}{m}=100$, below the threshold of
higher LLS formation. We see in Fig.~\ref{fig:III-2} the standard behaviour inherent to
the anisotropic mass model. The situation is very different in Fig.~\ref{fig:III-3},
where the slope $-\left( \frac{dH_{c2}^{{}}(90%
%TCIMACRO{\U{b0}}%
%BeginExpansion
{{}^\circ}%
%EndExpansion
)}{dT}\right) _{T=T_{c0}}=60\left( T/K\right) $ is well above the threshold.
At low temperature the angular dependence $H_{c2}^{{}}(\theta )$ clearly
reveals the transition between the higher LLS, making the overall shape of
the $H_{c2}^{{}}(\theta )$ curve very peculiar, and somewhat similar to the
corresponding results of section II.

Note that in isotropic systems the FFLO modulation vector $\mathbf{Q%
}$ is directed along the applied magnetic field. In anisotropic
superconductor the FFLO modulation is described by $\exp\left( iQ\left( \frac{%
\widetilde{H}_{z}}{\widetilde{H}}z'+\frac{\widetilde{H}_{x}}{\widetilde{H}%
}x\right) \right)$ $\sim \exp\left( iQ\left( \frac{H\cos \theta}{\widetilde{H}%
}\sqrt{\frac{m_{z}}{m}}z +\sqrt{\frac{m}{m_{z}}}\frac{H\sin\theta}{%
\widetilde{H}}x\right) \right)$, that is \newline
$\sim \exp\left( \frac {iQ}{\sqrt{\cos^{2}\theta+\frac{m}{m_{z}}\sin^{2}\theta%
}}\left( z\cos \theta^{{}}+\frac{m}{m_{z}}x\sin\theta\right) \right)$.
Therefore  the angle $\theta'$ that the direction of the FFLO modulation makes
with the $z$ axis is given by $\tan\theta'=\frac{m}{m_{z}}\tan\theta$. This means
that for quasi-2D anisotropy the FFLO modulation vector deviates from the
field direction toward the $z$ axis, while for the quasi-1D anisotropy it
lies closer to the $xy$ plane.

The appearence of the higher LLS in quasi-2D superconductors when the
magnetic field direction approach the $xy$ plane is consistent with the
prediction of such states in 2D superconductors in tilted magnetic field 
\cite{Bulaevskii74},\cite{BuzdinBrisonEuro},\cite{ShimaharaRainer97}.

\section{The origin of the first order superconducting transition in CeCoIn$_{5}$
 and high contrast vortex phase} \label{sec:4}

In a magnetic field the superconducting transition in CeCoIn$_{5}$ becomes
slightly first order below $0.3T_{co}$ for the field along the tetragonal $z$
axis and below $0.4T_{co}$ for the field in the $xy$ basal plane \cite{Izawa2}. 
The change of the transition order occurs at a magnetic field lower than that of
the presumed FFLO transition. Another particularity of this compound is the
field induced antiferromagnetic transition when the magnetic field is in the
basal plane \cite{Kenzelmann,Young}. This antiferromagnetic phase
exists only in the mixed state and basically in the same region, where the
FFLO state is expected. Neutron scattering experiments \cite{Kenzelmann}
reveal a small value of the magnetic moment on cerium sites $\sim 0.15\mu
_{B}$ oriented along the tetragonal axis. We may expect that the normal
state of CeCoIn$_{5}$ is very close to the magnetic instability of the
itinerant type. Indeed \ the measurements of the magnetic susceptibility
reveal its strong temperature increase at low temperature (at several $K$) 
\cite{Xiao}.

We propose to consider CeCoIn$_{5}$ as a system with two type of electrons,
one mostly localized on cerium sites and responsible for the magnetism and
the second strongly delocalized and responsible for superconducting
properties. In fact a multiband picture for CeCoIn$_{5}$ was already
discussed by several authors -- see for example~\cite{MatsudaShimahara07}.

In a magnetic field the electrons from the Ce band, which are polarized due
to the exchange interaction, will create, in addition to Zeeman field, some
internal field acting on the spins of the superconducting electrons.

We may describe this situation by a simple Ginzburg-Landau functional
introducing, in addition to the superconducting order parameter, the magnetic
moment $M$ of Ce sub-band:
\begin{equation}
F(M,\Psi )=-MH+A(T)M^{2}+\alpha \left( H+\gamma M-H_{p}(T)\right) \left\vert
\Psi \right\vert ^{2}+\frac{b}{2}\left\vert \Psi \right\vert ^{4}+\delta
F_{orb},  \label{F(M)}
\end{equation}
where $H_{p}(T)$ is the paramagnetic critical field and $\delta F_{orb}$
describes the contribution of the orbital effect. The constant $\gamma $
describes the contribution of polarized Ce band electrons to the field
acting on the spin of the superconducting electrons. Minimizing (\ref{F(M)})
over $M$ we find
\begin{equation}
M=\frac{H-\gamma\alpha\left\vert \Psi\right\vert ^{2}}{2A(T)}  \label{M(H)}
\end{equation}
and finally substituting this expression into (\ref{F(M)}) we obtain the pure
superconducting functional
\begin{equation}
\delta F_{s}(\Psi)=-\frac{H^{2}}{4A(T)}+\alpha\left( H+\gamma\frac{H}{2A(T)}%
-H_{p}(T)\right) \left\vert \Psi\right\vert ^{2}+\frac{1}{2}\left( b-\frac{%
\alpha^{2}\gamma^{2}}{4A(T)}\right) \left\vert \Psi\right\vert ^{4}.
\end{equation}

Here the role of the Ce band magnetization is the renormalization of the
Zeeman field $H\rightarrow H\left( 1+\frac{\gamma}{2A(T)}\right) $ and the shift of
coefficient $\ b$ of the  $\left\vert \Psi\right\vert ^{4}$ term $b\rightarrow b-%
\frac{\alpha^{2}\gamma^{2}}{4A(T)}.$ Whatever sign of the exchange
interaction $\gamma$, it decreases the coefficient of the $\left\vert \Psi\right\vert ^{4}$
term. With an increase of polarization at the normal state $M/H=%
\frac{1}{A(T)}$,  it may even become negative. This means that the superconducting
transition becomes first order. We believe that namely such a
situation is realized in CeCoIn$_{5}$ at low temperature, explaining the
observed first order transition below $\left( 0.3-0.4\right) T_{co}$.

Moreover the contribution to the magnetization from the Ce band (\ref{M(H)})
depends on the profile of the superconducting order parameter. In the
vortex state the maxima of the magnetization would be at the vortex core,
where the superconducting order parameter vanishes. This circumstance may
strongly increase the amplitude of the magnetic field modulation and then
explain the anomalous behavior of the form factor of the vortex structure
observed in CeCoIn$_{5}$ at low temperature \cite{Bianchi,White}.

Previously Ichioka and Machida \cite{Ichio} performed an extensive
numerical analysis of the role of the Pauli paramagnetic effect in the
context of quasi-classical Eilenberger theory. They demonstrated the
modulation of electron spin susceptibility in the mixed state, increasing
the electron magnetization in the vortex core and thus they have explained
the anomalous behaviour of the form factor. In the recent paper \cite{Michal},
a similar problem has been treated by a variational method. Note that in
these theories, only one electron band is implied, and therefore it is
difficult to explain the occurrence of the first order superconducting
transition. In some sense, our model provides an additional mechanism for the
increase of the contrast of the magnetic field modulation. If we consider that
the Ce band magnetism contributes to the temperature dependent
susceptibility, while the band responsible for superconductivity gives
the temperature independent contribution, then from the experimental data 
\cite{Xiao} we may roughly estimate that they are equally involved at low
temperature ($T\sim 2K).$ \bigskip

\section{Conclusions}

In real compounds, the crystal structure plays a  dominant role
in determining the type of  FFLO state. Of course it will also influence
the vortex structure. The FFLO state may be characterised by an
uni-dimensional modulation of the order parameter, and/or by the emergence
of higher Landau level states. This is a crucial difference with
superconductivity without FFLO state, where the crystal structure influence
only on the type of the Abrikosov vortex lattice. The higher Landau level
FFLO states should be realized in systems with strong uni-axial anisotropy and
near in-plane orientation of the magnetic field. In such a case, the higher
LL states should lead to an unusual angular dependence of $H_{c2}$. Finally 
we propose a simple explanation of the first order transition at low
temperature in CeCoIn$_{5}$ based on the Ce-band magnetic
contribution.

\appendix

\section{Modified Ginzburg-Landau theory}

In the paramagnetic limit the MGL functional quadratic over $\Psi$ is 
\begin{equation}
\mathcal{F}= \Psi^* \Big( \alpha - \sum_j g_j \Pi_j^2 + \hat{L}_4(\Pi_j) %
\Big)\Psi,
\end{equation}
where $\alpha = \alpha_0(T-T_{cu}(H))$, $\Pi_j= -i \partial_j + 2\pi A_j
/\Phi_0$ (with $j=x,y,z$) and $\hat{L}_4(\Pi_j)$ is the forth-order part of
the $\Pi_j$ expansion.

\subsection{Rescaling of z coordinate}

\begin{figure}[tbp]
\begin{center}
\scalebox{.6}{
 \includegraphics*{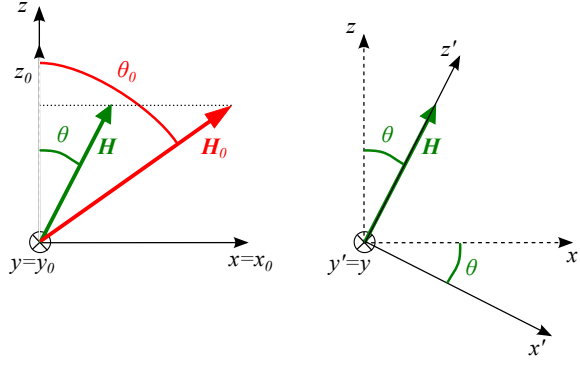}}
\end{center}
\caption{Successive changes of coordinates: a scaling of the $z$-coordinate
followed by a rotation around the $y$-axis.}
\label{fig:frame}
\end{figure}

The functional is invarient by the tranformations of the crystal symmetry
group. In the tetragonal symmetry $g_x=g_y =g\neq g_z$ so, to recover an
isotropic expression of the second-order part, one can rescale the $z$%
-coordinate as $z=z_0 \sqrt{m_z/m_x}$ and the vector potential as $\mathbf{A}%
= (A_{0x},A_{0y},\sqrt{m_x/m_z} A_{0z})$. Hence $\Pi_{z_0}= \sqrt{m_z/m_x}
\Pi_z$ so that $\sum_j g_j \Pi^2_{j0} = g \sum_j \Pi_j^2$ (since $g_j
\propto m_j^{-1}$). In doing so the magnetic field is transformed as 
\begin{equation}
\mathbf{H}=\left( \sqrt{\frac{m_x}{m_z}} H_{0x}, \sqrt{\frac{m_x}{m_z}}
H_{0y}, H_{0z} \right),
\end{equation}
where $\mathbf{H}=\nabla_{\mathbf{r}} \times \mathbf{A}$ and $\mathbf{H}%
_0=\nabla_{\mathbf{r}_0} \times \mathbf{A}_0$. The angles that $\mathbf{H}$
and $\mathbf{H}_0$ have with the $z$-axis (see Fig.~\ref{fig:frame}) are
then related by the equality 
\begin{equation}
\theta = \arctan \left( \sqrt{\frac{m_x}{m_z}} \tan{\theta_0} \right).
\end{equation}
Hence, after rescaling, the functional in the tetragonal symmetry has the
general expression 
\begin{equation}
\mathcal{F} = \Psi^* \bigg[ \alpha - g \sum_j{\Pi_j^2} + \gamma \Big( \sum_j{%
\Pi_j^2} \Big)^2 + \varepsilon_z \Pi_z^4 + \frac{\varepsilon_x}{2} \left\{
\Pi_x^2, \Pi_y^2 \right\} + \frac{\tilde{\varepsilon}}{2} \Big( \{ \Pi_x^2,
\Pi_z^2 \} + \{ \Pi_y^2, \Pi_z^2 \} \Big) \bigg] \Psi,
\end{equation}
where the anti-commutator $\{O_1,O_2\}\equiv O_1 O_2+ O_2 O_1$. Note that in
order to recover the functional used in previous work~\cite{Denisov}, the
term $(2\pi/\Phi_0)^2 \big[  \tilde{\varepsilon} (H_x^2 + H_y^2) +
\varepsilon_x H_z^2 \big]$ must be added in our expression. The latter term
only shifts the energy by a constant so the solution for the order parameter
is not modified.

\subsection{Expansion over Landau levels for the field applied in the $xz$
plane}

In order to determine the transition temperature, one needs to find the
eigenvalues of the operator $\hat{L}$ which is a polynomial of $\Pi_j$. When
the magnetic field is in the $xz$-plane, it is convenient to work in the
rotated coordinate frame $(x^{\prime},y^{\prime},z^{\prime})$ where the $%
z^{\prime}$ axis points in the same direction as $\mathbf{H}$ (see Fig.~\ref%
{fig:frame}). By the change of coordinates the gradient operators are
transformed as $\Pi_x = \cos{\theta} \Pi_{x^{\prime}} + \sin{\theta}
\Pi_{z^{\prime}}$, $\Pi_y=\Pi_{y^{\prime}}$, and $\Pi_z = -\sin{\theta}
\Pi_{x^{\prime}} + \cos{\theta} \Pi_{z^{\prime}}$. Since the field $\mathbf{H%
}$ is along the $z^{\prime}$ axis, the operator $\Pi_{z^{\prime}}$ commutes
with both $\Pi_{x^{\prime}}$ and $\Pi_{y^{\prime}}$. So, with an adequate
choice of gauge, the eigenfunctions of $\hat{L}$ can be looked for in the
form $\Psi= \exp(i q_z z^{\prime}) \phi(x^{\prime},y^{\prime})$ where $q_z$
is the FFLO modulation vector along the field direction. In the abscence of
the anisotropic forth-order terms, $\phi$ is a Landau level. Functions $\varphi_{q_z,n}\equiv \exp(i q_z z^{\prime})\varphi_n(x^{\prime},y^{\prime})$, where $\varphi_n$ are Landau levels, then
form a natural basis over which to expand the solution in the anisotropic
case. We use the orthonormal basis set composed by the states $\varphi_{2n}
\equiv \left( \eta^{\dagger} \right)^{2n} \varphi_0 / \sqrt{(2n)!}$ and $%
\varphi_{2n+1} \equiv -i \left( \eta^{\dagger} \right)^{2n+1} \varphi_0 / 
\sqrt{(2n+1)!}$. Here $\eta^{\dagger}$ is the operator of Landau-level
creation defined as $\eta^{\dagger} \equiv \frac{\xi_H}{\sqrt{2}}
(\Pi_{y^{\prime}} - i \Pi_{x^{\prime}} )$ where the magnetic length 
\begin{equation}
\xi_H \equiv \sqrt{\frac{\Phi_0}{2\pi H}},
\end{equation}
and $\varphi_0$ is the normalized lowest Landau level defined by $\eta
\varphi_0 = 0$. With $\eta = \frac{\xi_H}{\sqrt{2}} (\Pi_{y^{\prime}} + i
\Pi_{x^{\prime}} )$, one can easily check for example that $\eta
\eta^{\dagger} - \eta^{\dagger} \eta =1$ and $\Pi_{x^{\prime}}^2 +
\Pi_{y^{\prime}}^2 = \xi_H^{-2} ( 2\eta^{\dagger} \eta +1)$.

After expressing the operator $\hat{L}$ as a function of $\eta$ and $%
\eta^{\dagger}$, the matrix elements $L_{m,n}\equiv \int \varphi_{q_z,m}^* 
\hat{L} \varphi_{q_z,n}$ are found as 
\begin{equation}
L_{m,n} = \frac{\gamma}{\xi_H^4} \Big[ \big((2n+1+k^2-k_0^2)^2 -k_0^4 \big) %
\delta_{m,n} + L^{(\varepsilon)}_{m,n} \Big]  \label{eq:Lmn}
\end{equation}
with 
\begin{equation}
k\equiv\xi_H q_z \;\;\mathrm{and} \;\; k_0\equiv\xi_H \sqrt{\frac{g}{2\gamma}%
}.
\end{equation}
They connect states which are separated by at most four levels. Within the
above choice of basis set, the matrix is real symmetric 9-diagonal and the
non-zero terms above the diagonal are given by the anisotropic contributions 
\begin{eqnarray}
L^{(\varepsilon)}_{n,n} & = & \frac{\varepsilon_z}{4\gamma} \Big( 4 c^4 k^4
+ 12 c^2 s^2 (2n+1)k^2 + 3 s^4 (2n^2+2n+1)\Big)  \notag \\
& + & \frac{\tilde{\varepsilon}}{4\gamma} \Big( s^2 (3c^2-1) + 2 s^2
(1+3c^2) (n^2 +n) + 2(1+c^2 - 6 c^2s^2)(2n+1)k^2 + 4 c^2 s^2 k^4 \Big) 
\notag \\
& + & \frac{\varepsilon_x}{4\gamma} \Big( c^2 (2n^2+2n-1) + 2 s^2 (2n+1)k^2 %
\Big),  \notag \\
L^{(\varepsilon)}_{n,n+1} & = & (-1)^n \sqrt{n+1} k \sin{\!2\theta} \bigg[ 
\frac{\varepsilon_z}{\gamma} \Big( \frac{3}{\sqrt{2}} s^2 (n+1) + \sqrt{2}
c^2 k^2 \Big) - \frac{\varepsilon_x (n+1)}{2 \sqrt{2} \gamma}  \notag \\
& + & \frac{\tilde{\varepsilon}}{\sqrt{2}\gamma} \Big( \frac{1+3 \cos{2\theta%
}}{2} (n+1) - k^2 \cos{2\theta} \Big) \bigg],  \notag \\
L^{(\varepsilon)}_{n,n+2} & = & \sqrt{(n+2)!/n!} s^2 \bigg[ \frac{%
\varepsilon_x}{2\gamma} k^2 - \frac{\varepsilon_z}{\gamma} \bigg( \Big( n+%
\frac{3}{2} \Big) s^2 + 3 c^2 k^2 \bigg) + \frac{\tilde{\varepsilon}}{2\gamma%
} \Big( -c^2 (2n+3) + (6c^2-1) k^2 \Big) \bigg],  \notag \\
L^{(\varepsilon)}_{n,n+3} & = & (-1)^n \sqrt{(n+3)!/n!} \frac{k\sin{\!2\theta%
}}{\sqrt{2}} \bigg[ (\frac{\tilde{\varepsilon}}{\gamma} - \frac{\varepsilon_z%
}{\gamma}) s^2 - \frac{\varepsilon_x}{2\gamma} \bigg],  \notag \\
L^{(\varepsilon)}_{n,n+4} & = & \sqrt{(n+4)!/n!} \frac{1}{4} \bigg[ (\frac{%
\varepsilon_z}{\gamma} - \frac{\tilde{\varepsilon}}{\gamma}) s^4 - \frac{%
\varepsilon_x}{\gamma} c^2 \bigg],
\end{eqnarray}
with $s=\sin{\theta}$ and $c=\cos{\theta}$.

\begin{acknowledgments}
The authors are grateful to A. Melnikov and A. Samokhvalov for useful discussions and comments. This work was supported by the French ANR project SINUS and European IRSES program SIMTECH (Contract No. 246937).
\end{acknowledgments}

\end{document}